\documentclass[11pt,twoside]{article}

\usepackage{asp2006}
\usepackage{epsf}
\usepackage{lscape}
\usepackage{graphicx}
\usepackage{url}

\begin{document}

\title{VLTI\,/\,MIDI Observations of the Massive Protostellar Candidate NGC\,3603 IRS\,9A}

\author{Stefan Vehoff,\altaffilmark{1,2} Dieter E.~A. N{\"u}rnberger,\altaffilmark{1}
Christian A. Hummel,\altaffilmark{3} and Wolfgang J. Duschl\altaffilmark{4,5}}

\altaffiltext{1}{European Southern Observatory, Alonso de C\'ordova 3107, Casilla 19001,
Vitacura, Santiago, Chile; svehoff@eso.org}
\altaffiltext{2}{Zentrum f{\"u}r Astronomie, Institut f{\"u}r Theoretische Astrophysik,
Albert-Ueberle-Stra{\ss}e 2, 69120 Heidelberg, Germany}
\altaffiltext{3}{European Southern Observatory, Karl-Schwarzschild-Stra{\ss}e 2, 85748
Garching, Germany}
\altaffiltext{4}{Institut f{\"u}r Theoretische Physik und Astrophysik, Universit{\"a}t
Kiel, Leibnizstra{\ss}e 15, 24118 Kiel, Germany}
\altaffiltext{5}{Steward Observatory, The University of Arizona, Tucson, AZ 85721, USA}

\begin{abstract}
  We used MIDI, the mid-infrared interferometric instrument of the VLTI, to observe
  the massive protostellar candidate IRS 9A, located at a distance of about 7 kpc at
  the periphery of the NGC 3603 star cluster. Our ongoing analysis shows that MIDI
  almost fully resolves the object on all observed baselines, yet below 9\,$\mu$m we
  detect a steep rise of the visibility. This feature is modelled as a combination of
  a compact hot component and a resolved warm envelope which lowers the correlated
  flux at longer wavelengths. The extended envelope can already be seen in both MIDI's
  acquisition images and in complementary data from aperture masking observations at
  the Gemini South telescope. Its shape is asymmetric, which could indicate a circumstellar
  disk inclined against the line of sight. The compact component is possibly related to
  the inner edge of this (accretion) disk. The uncorrelated mid-infrared spectrum appears
  featureless and could be caused by optically thick emission without a significant
  contribution from the disk atmosphere.
\end{abstract}

\section{Introduction}
The giant H\,II region NGC 3603 is powered by one of the densest clusters of high mass
stars known in our galaxy. Due to its vicinity to this cluster, IRS 9A has been liberated
from most of the gas and dust of its natal molecular cloud by the strong stellar winds
and ionising radiation arising from the nearby O-type main-sequence stars. This offers
the unique possibility to observe a high mass star at infrared wavelengths during its
relatively early evolutionary phase. 

\section{Observations and data reduction}
\begin{figure}[t]\center
  \begin{minipage}{8.8cm}
    \small
    \begin{tabular}{ccccc}
      \tableline
      \noalign{\smallskip}
      Telescopes & Date       & Time     & B$_{\mbox{\scriptsize p}}$ [m] & P.A. [$\deg$] \\
      \noalign{\smallskip}
      \tableline
      \noalign{\smallskip}
      UT2--UT3   & 2005-02-27 & 08:35:10 &               31.8             & 79.8  \\
         ''      & 2005-02-28 & 06:48:35 &               38.0             & 60.8  \\
         ''      &      ''    & 07:43:50 &               34.9             & 70.8  \\
         ''      &      ''    & 08:33:25 &               31.7             & 80.2  \\
         ''      &      ''    & 09:21:12 &               28.1             & 90.1  \\
      UT3--UT4   & 2005-03-03 & 07:10:49 &               62.5             & 131.9 \\
         ''      &      ''    & 08:54:33 &               62.0             & 155.5 \\
      \noalign{\smallskip}
      \tableline
    \end{tabular}
  \end{minipage}
  \hfill
  \begin{minipage}{4.3cm}
    \includegraphics[width=4.3cm]{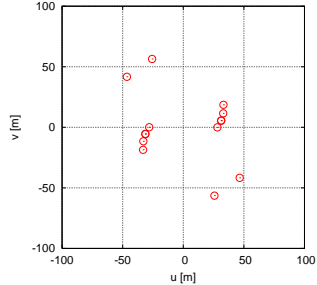}
  \end{minipage}
  \begin{minipage}{13.4cm}
    \caption{Log of observations and plot of the {\it u,v} - coverage of IRS 9A} \label{f_log}
  \end{minipage}
\end{figure}

IRS 9A was observed with MIDI \citep{svehoff_leinert2003} during three nights in 2005.
The first two observations were carried out on the 26th and the 27th of February, using
the VLT telescopes UT2 and UT3. The third observation, performed on the 2nd of March, was
executed with telescopes UT3 and UT4. A short observing log, containing the length of the
projected baselines and the position angles, and a plot of the corresponding {\it u,v} -
coverage are shown in Figure \ref{f_log}.

The star HD107446 has been chosen as a calibrator and was observed repeatedly during the
three aforementioned nights. The rather complex observing procedure is described in detail
in \citet{svehoff_przygodda2003} and \citet{svehoff_leinert2004}, and will not be presented
here. Data reduction was performed with the MIA+EWS package, which is provided by the MIDI
consortium.

\subsection{Images}
\begin{figure}[t]\center
  \begin{minipage}{5cm}
    \includegraphics[width=5cm]{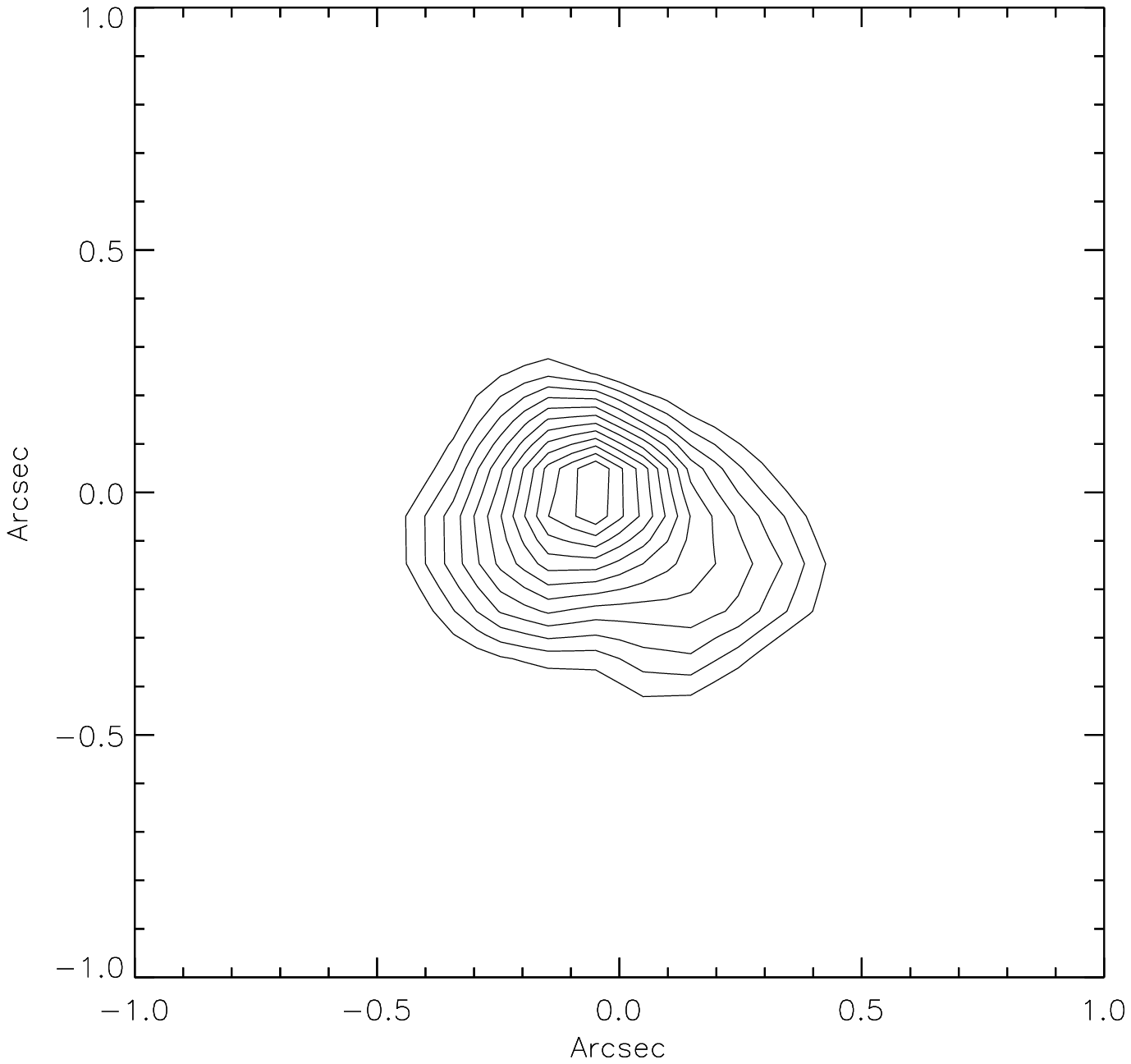}
  \end{minipage}
  \hfill
  \begin{minipage}{5cm}
    \includegraphics[width=5cm]{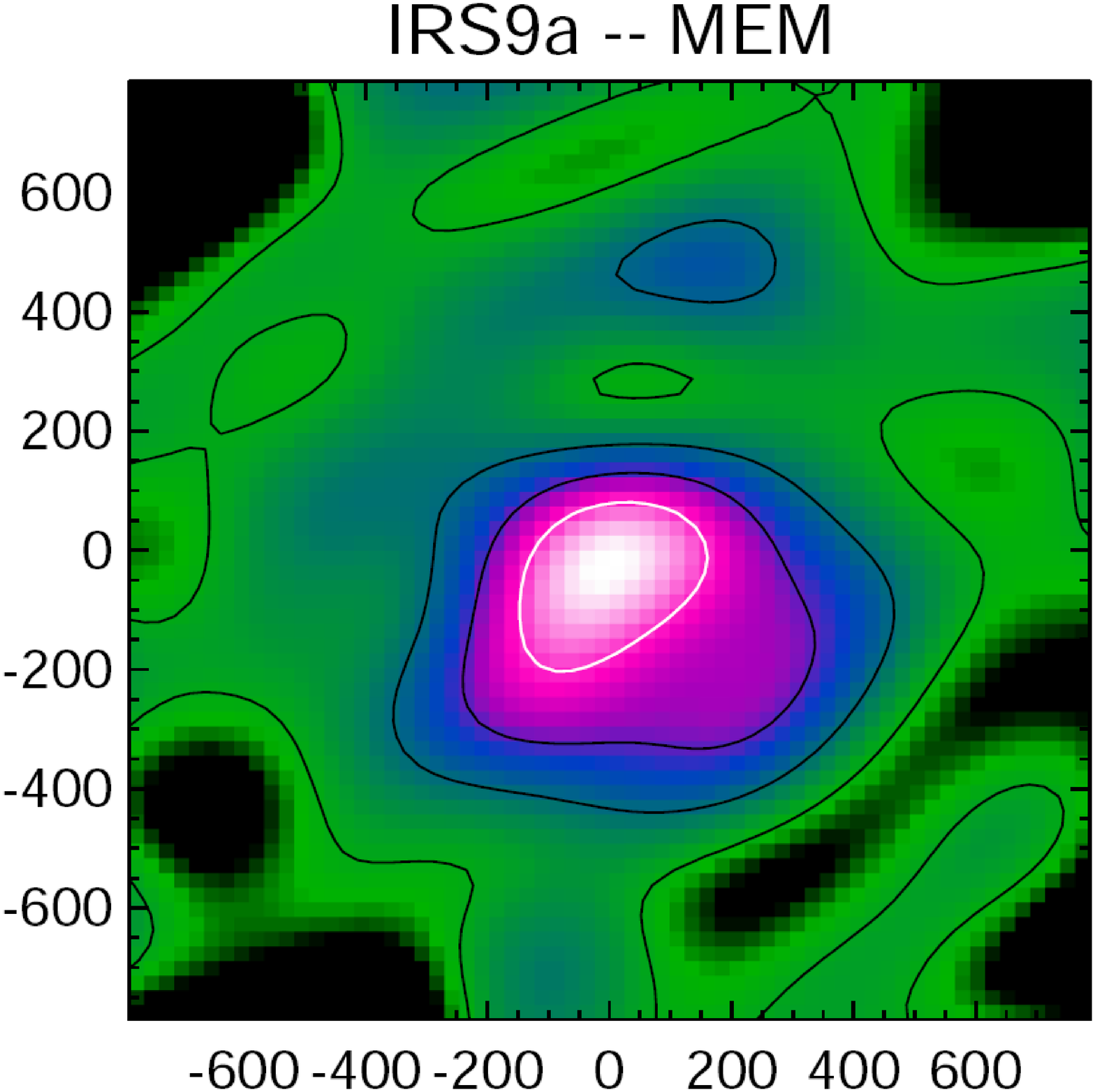}
  \end{minipage}
  \begin{minipage}{13.4cm}
    \caption{Left: Acquisition image from VLTI\,+\,MIDI. Right: MEM \mbox{image} from
      Gemini South\,+\,T-ReCS aperture masking data (J. Monnier, priv. comm.; scale on
      both axes is in units of mas).} \label{f_images}
  \end{minipage}
\end{figure}

Before any interferometric data are taken, MIDI records acquisition images to center
the target in the FOV. Since the scaling of the Fried parameter $r_0$ is favourable
for observations in the infrared ($r_0 \propto \lambda^{6/5}$) and an adaptive optics
system is used, these images are diffraction limited and have a resolution of about
$ 0 \farcs 3 $. They show that the envelope of IRS 9A is already partly resolved by
a single 8\,m telescope in the N band, and that the shape of the extended flux is
asymmetric. This asymmetry is further supported by aperture masking data from T-ReCS
on Gemini South, which was kindly provided by John D. Monnier (priv. comm.). Figure
\ref{f_images} shows one of the acquisition images taken with MIDI and the image
from Gemini South, reconstructed with the Maximum Entropy Method. They agree in
overall shape and show that the circumstellar emission is indeed quite extended. At
the distance of IRS 9A, $ 0 \farcs 1 $ correspond to 700\,AU, and the bulk of the
emission therefore originates in a region of about 3000\,AU in diameter.

\subsection{Spectrum}
The mid-infrared spectrum is obtained by standard chopping techniques and does not show
any prominent emission or absorption features (see Figure \ref{f_spec_vis}). A small dip
and bump at approximately 9.4\,$\mu$m can be seen in the spectra from the first two
nights, but this is most likely an artefact of imperfect calibration, since there is
strong absorption due to Ozone in the atmosphere at this wavelength. On top of that, one
apparently sees a faint silicate absorption feature ranging from 9\,$\mu$m to 11.5\,$\mu$m.
Nevertheless the overall spectrum can be reproduced by a single black body of about 250 K.

\subsection{Visbilities}
\begin{figure}[t]\center
  \begin{minipage}{6cm}
    \includegraphics[width=6cm]{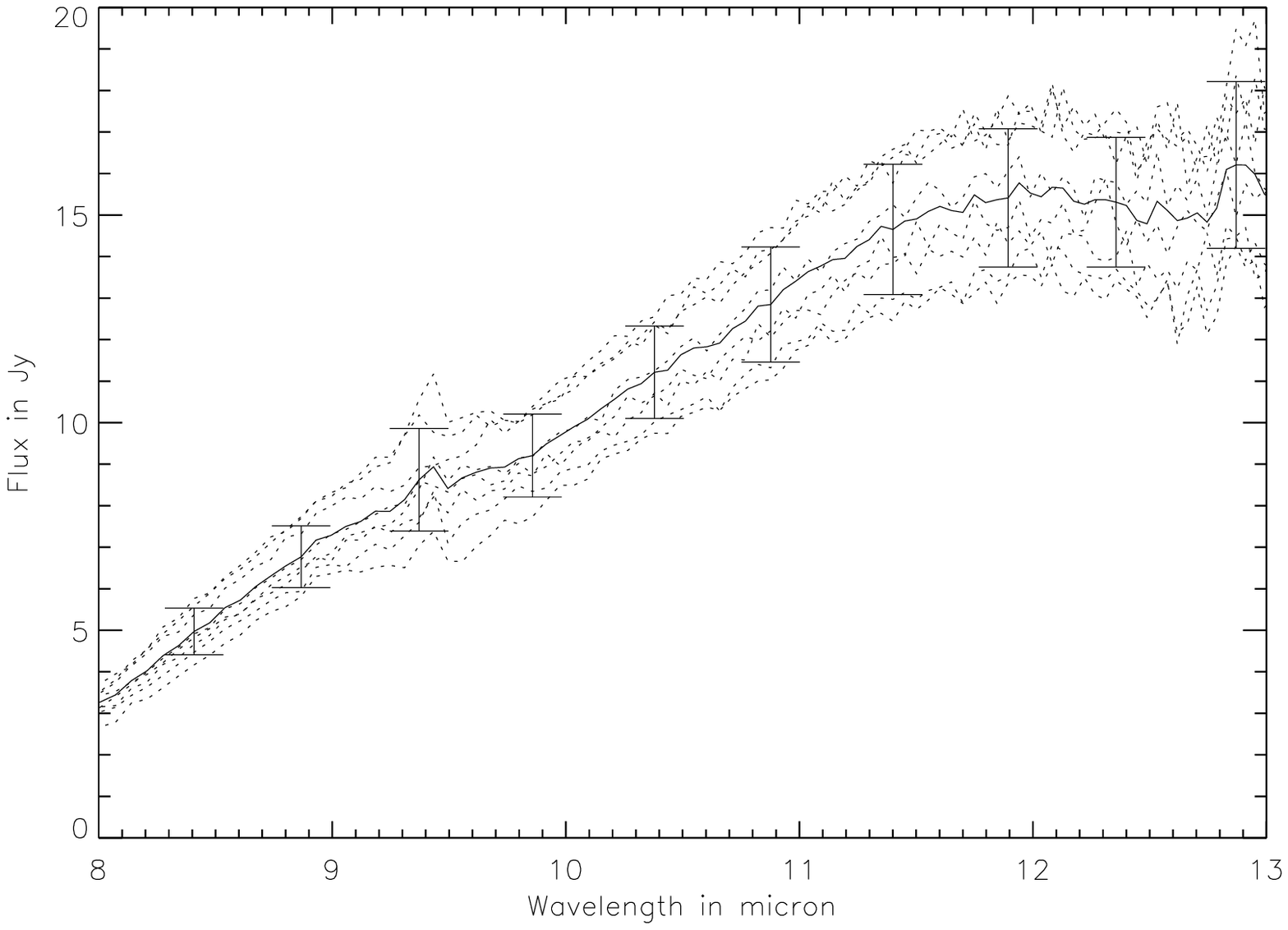}
  \end{minipage}
  \hfill
  \begin{minipage}{6cm}
    \includegraphics[width=6cm]{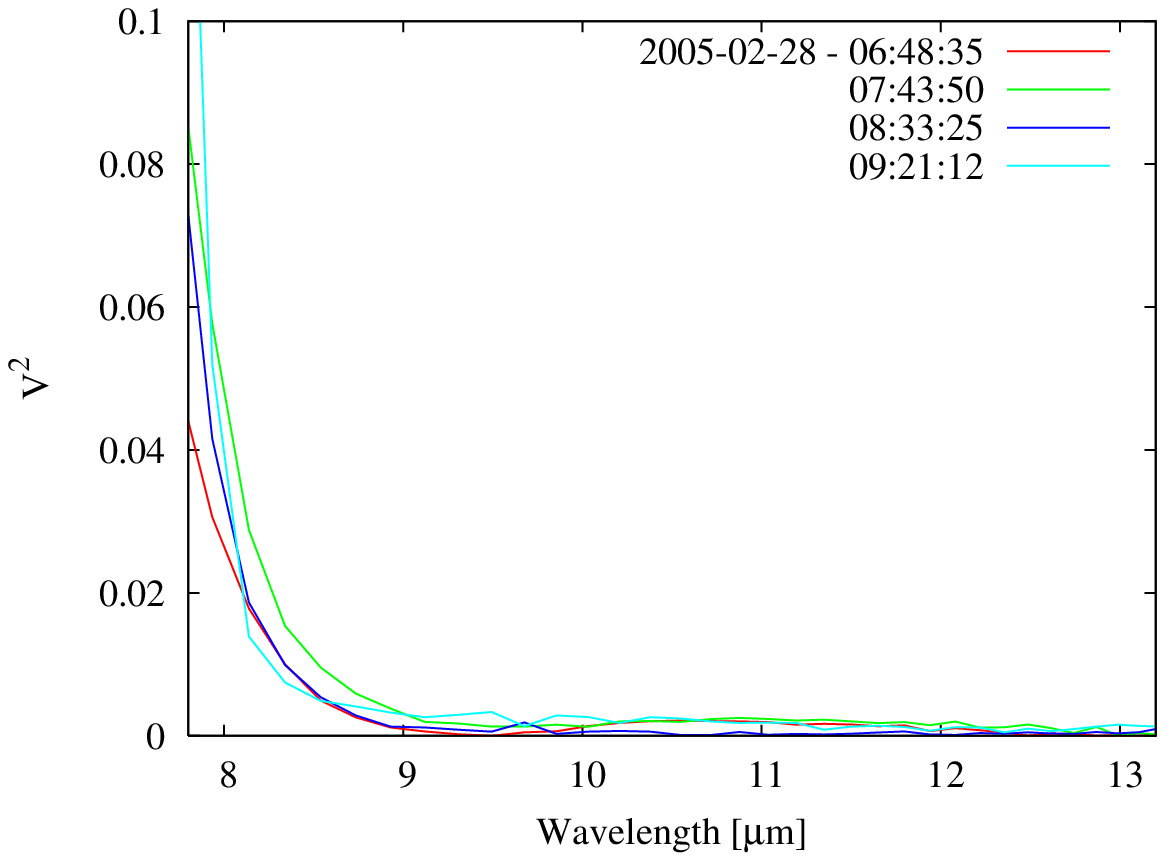}
  \end{minipage}
  \begin{minipage}{13.4cm}
    \caption{Left: Spectra obtained during the second night. The dashed lines are
      the individual measurements, the solid line shows the average together with
      the standard deviation as error bars. Right: Corresponding visibility
      measurements from the same night.} \label{f_spec_vis}
  \end{minipage}
\end{figure}

Given the detection of the extended envelope around IRS 9A we already expected the
visibilities to be quite low. Figure \ref{f_spec_vis} shows the calibrated and squared
visibilities of the second night. Their value is basically zero above a wavelength of
9\,$\mu$m, but there is a very steep rise with decreasing wavelength below 9\,$\mu$m.
However, even the largest values are only of the order of 0.1 and lie at the very edge
of the atmospheric window. Except for one observation on the 3rd of March, all the
measurements are in good agreement and do not show a strong dependency on the position
angle, if any.

\section{Modelling the SED and the visibilities}
Apart from the spectrum taken with MIDI we have at hand additional mid-infrared fluxes
of IRS 9A \citep{svehoff_nuernberger2003}. We use two publicly available radiative
transfer codes to model the spectral energy distribution of IRS 9A: The first one is
DUSTY \citep{svehoff_ivezic1999}, which solves the radiation transport for a spherical
dust distribution. The second one is MC3D \citep{svehoff_wolf1999}, which can handle
more complex geometries but uses Monte Carlo methods to solve the radiation transport.
Both models are able to reproduce the SED quite well.

\begin{figure}[t]\center
  \begin{minipage}{6cm}
    \includegraphics[width=6cm]{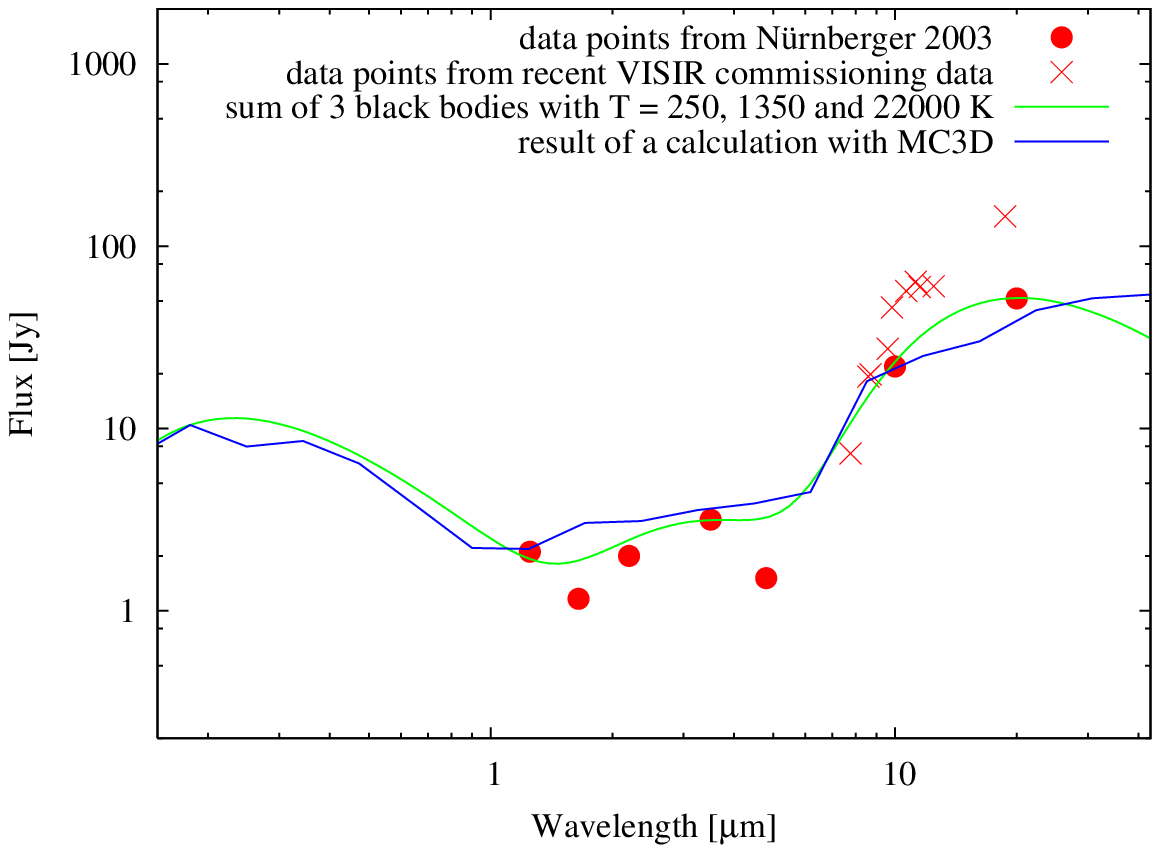}
  \end{minipage}
  \hfill
  \begin{minipage}{6cm}
    \includegraphics[width=6cm]{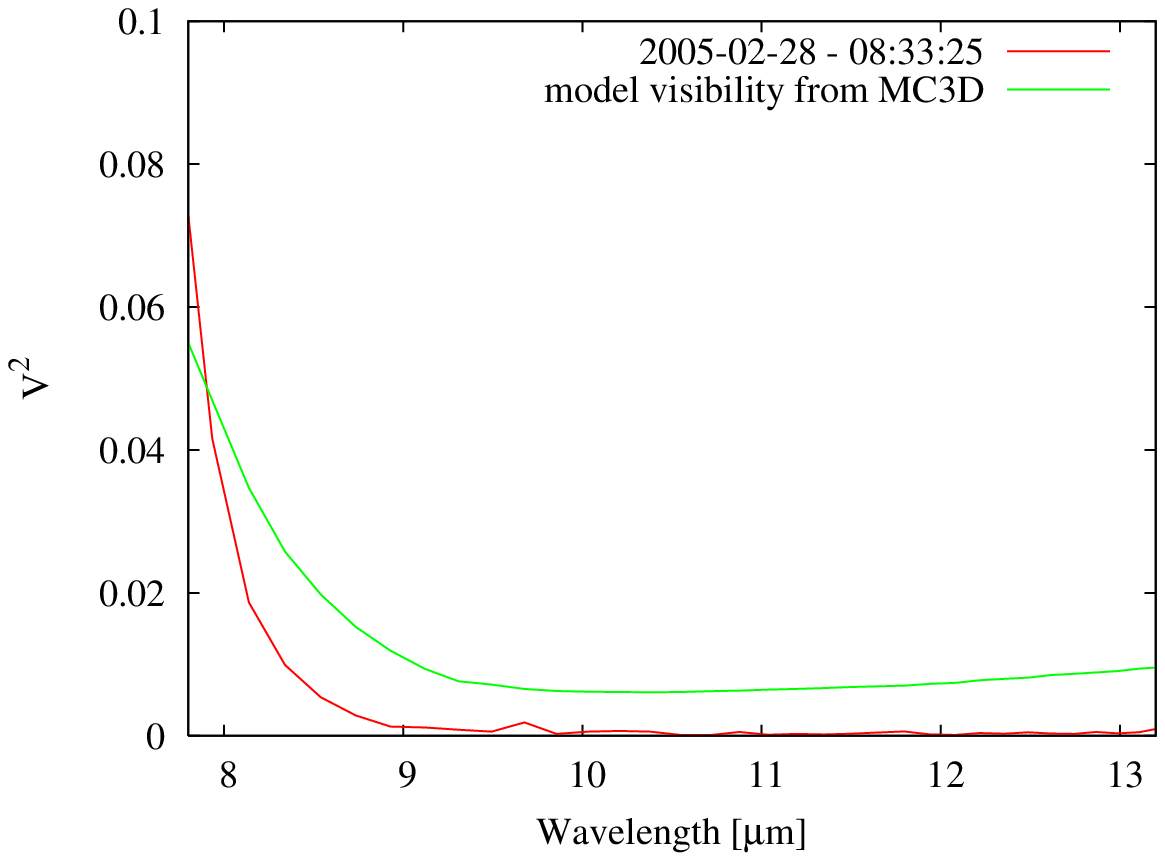}
  \end{minipage}
  \begin{minipage}{13.4cm}
    \caption{Left: SED model obtained with MC3D in comparison to the sum of three black
      bodies and our additional near-infrared data. Right: The visibility of this model
      compared to one of the measurements.} \label{f_models}
  \end{minipage}
\end{figure}

The radiative transfer codes allow to create maps of the surface brightness for a
given wavelength, which can then be used to calculate the wavelength dependent
visibilities. The DUSTY model yields visibilities which are not too far from our
measurements, but the visibility below 9\,$\mu$m rises rather gently, and it also
shows a bump between 9 and 10\,$\mu$m. The visibilities from our MC3D model produce
a better fit to the observed data, but the slope below 9\,$\mu$m is still not steep
enough. Figure \ref{f_models} shows the results obtained with MC3D.

\section{Discussion}
Our ongoing analysis of the MIDI data shows IRS 9A to be almost completely resolved
on baselines of the order of 30\,m. The mid-infrared spectrum is featureless, and most
likely caused by emission from a warm envelope. The models are so far unable to account
for the steep rise of the visibility towards the short wavelength end, so that the
nature of this compact component remains unclear and further refinements of the models
are needed.

\acknowledgements We would like to thank J.~D. Monnier for kindly providing his IRS 9A
data from Gemini South.

\end{document}